\documentclass[sigconf,nonacm,screen]{acmart}

\AtBeginDocument{%
  \providecommand\BibTeX{{%
    \normalfont B\kern-0.5em{\scshape i\kern-0.25em b}\kern-0.8em\TeX}}}

\setcopyright{none}

\usepackage{tikz}  
\usepackage{listings}  
\usepackage{siunitx}  
\usepackage{algorithm2e}  

\graphicspath{ {images/} }

\begin{document}

\title{Parallel Oculomotor Plant Mathematical Model for Large Scale Eye Movement Simulation}

\author{Alex Karpov}
\email{ak26@txstate.edu}
\author{Jacob Liberman}
\email{jl1496@txstate.edu}
\author{Dillon Lohr}
\email{djl70@txstate.edu}
\author{Oleg Komogortsev}
\email{ok11@txstate.edu}
\affiliation{%
  \institution{Texas State University}
  \streetaddress{601 University Drive}
  \city{San Marcos}
  \state{Texas}
  \postcode{78666}
}









\begin{abstract}
The usage of eye tracking sensors is expected to grow in virtual (VR) and augmented reality (AR) platforms.
Provided that users of these platforms consent to employing captured eye movement signals for authentication and health assessment, it becomes important to estimate oculomotor plant and brain function characteristics in real time.
This paper shows a path toward that goal by presenting a parallel processing architecture capable of estimating oculomotor plant characteristics and comparing its performance to a single-threaded implementation.
Results show that the parallel implementation improves the speed, accuracy, and throughput of oculomotor plant characteristic estimation versus the original serial version for both large-scale and real-time simulation.
\end{abstract}





\maketitle

\section{Introduction}
Virtual (VR) and augmented reality (AR) technologies are being adopted at an increasing rate.
The agency Digi Capital has estimated that the global market for both technologies will reach \$150 billion in 2020, including \$120 billion for AR alone \cite{komogortsev2016nsf}.
Eye tracking is expected to become an intrinsic component of such devices.
An eye tracking based technology called foveated rendering---or perceptual compression---minimizes the computational burden of VR and AR \cite{komogortsev2004predictive}.

Given that the use and accuracy of eye tracking sensors in VR and AR are expected to grow, one should consider how they can be applied to areas beyond interacting with the environment \cite{koh2009input,komogortsev2007kalman}.
Eye tracking sensors can potentially provide a platform for other important applications including health assessment (e.g. concussion diagnosis) \cite{gobert2012automated} and user authentication via eye movements \cite{komogortsev2014tbi}.
Both of these applications require the ability to estimate physiological characteristics related to oculomotor plant and brain function in near real-time.
   
Oculomotor plant characteristics (OPCs) quantitatively describe the physiological and neurological components of a subject's oculomotor plant.
These include values for the eye globe and its surrounding tissues, ligaments, the six extraocular muscles each containing thin and thick filaments, tendon-like components, and the various tissues and liquids surrounding the eye globe \cite{wilkie1956muscle}.
By estimating a subject's unique OPC vector, eye tracking can be applied to biologically mediated applications such as health assessment and biometric identification.

Previously, a single-threaded architecture was proposed for OPC estimation and tested for user authentication \cite{komogortsev2008eye}.
One of the biggest drawbacks of that architecture was the slow speed at which it was able to estimate OPCs.
It took approximately 15 minutes to process a single eye movement on a personal computer \cite{komogortsev2014tbi}.
This simulation speed is unacceptable for either real-time authentication or health assessment.
This paper presents and evaluates a parallel architecture that extracts OPCs in real-time.
The architecture was tested with an implementation in CUDA, a framework that extends the C/C++ programming languages with extensions for parallel processing.
The manuscript begins by describing the human oculomotor system and the model architecture.
Next, it introduces the parallel model's structure and algorithms.
Then, it shares the results of a study to evaluate the model's fitness for large-scale and real-time eye movement simulation.
The paper concludes with directions for future work.


\section{Oculomotor plant mathematical model}
The oculomotor system comprises both mechanical and neurological components that control eye movement.
The \textit{oculomotor plant} includes the mechanical components of the oculomotor system: the eye globe, the six extraocular muscles that rotate the globe, and the tendons, tissue, and fluids that surround and connect to the eye.
The \textit{neuronal control signal} encompasses the neurological components of the oculomotor system.
These components cause the extraocular muscles to expand and contract.
The oculomotor plant responds to the neuronal control signal with mechanical actions that dictate the type, magnitude, and direction of the eye movement.

The oculomotor plant mathematical model (OPMM) is a system of equations that can reproduce characteristics of human eye movement such as amplitude, duration, and velocity.
The OPMM takes recorded eye movement data as input and generates an OPC vector as output.
The OPC vector can reproduce the characteristics of the eye movement (velocity, amplitude, duration, etc.) when fed into the simulator.
The OPMM is biologically inspired.
Its output vector values describe the hidden properties of the subject's oculomotor system, accounting for all of the major physiological components of the oculomotor plant.
Given the appropriate input parameters, the OPMM is capable of reproducing various types of human eye movements.
However, the OPMM is primarily concerned with saccades---rapid ballistic and stereotypical eye movements that occur between fixation points. 

The OPMM has been productively applied to a wide range of research areas, including diagnosing mild traumatic brain injuries \cite{gobert2012automated,komogortsev2009instantaneous}, biometric identification via eye movement \cite{komogortsev2014tbi}, evaluating eye gaze guided computer interfaces \cite{koh2009input,komogortsev2007kalman,komogortsev2009instantaneous}, and predicting eye movements for eye gaze guided compression and foveated rendering \cite{komogortsev2004predictive}.

\subsection{Oculomotor plant characteristics}
The OPC parameter vectors map to the subject's hidden biological characteristics.
The OPCs quantitatively describe the physical and neurological properties exhibited by a subject's oculomotor system.
OPC parameters that represent physical properties include:
\begin{enumerate}
    \item \textit{Series elasticity}: the resistive force of a tendon in an extraocular muscle as it expands or contracts.
    \item \textit{Length-tension}: the relationship between the length of an extraocular muscle and the force it exerts. This relationship exists due to properties of the filaments inside of a muscle.
    \item \textit{Force-velocity}: the relationship between the velocity of muscle expansion/contraction and the force it exerts. This phenomenon exists due to the rate of the chemical reactions that occur in the extraocular muscle.
    \item \textit{Tension slope} and \textit{tension intercept}: properties of the extraocular muscle force that is exerted at various eccentricities at the equilibrium points of the oculomotor plant.
    \item \textit{Inertial mass} of the eye globe.
    \item \textit{Passive viscosity} of the tissue surrounding the eye globe.
\end{enumerate}

\textit{Agonist} muscles initiate the eye movement in a particular direction when innervated.
\textit{Antagonist} muscles resist the force generated by agonist muscles in order to decelerate or stabilize the eye globe.
Series elasticity, the length-tension relationship, the force-velocity relationship, and tension slope each vary between the agonist and antagonist muscles during eye movement.
Passive viscosity, the tension intercept during fixation, and the inertial mass of the eye globe are constant.

OPC parameters also represent the neuronal control signal, which is modeled as a pulse-step function.
The step indicates the magnitude of the neuronal control signal during a fixation.
The pulse indicates the magnitude of the signal during a saccade.
OPC parameters related to the neuronal control signal include:
\begin{enumerate}
    \item \textit{Activation time}: latency associated with propagating the signal at movement onset.
    \item \textit{Deactivation time}: latency associated with propagating the signal at movement offset.
    \item \textit{Neural pulse height}: magnitude of force generated by musculature during innervation.
    \item \textit{Neural pulse width}: duration of the neural pulse.
\end{enumerate}

Activation/deactivation time are both constants and neural pulse height varies across agonist and antagonist muscles.

\subsection{Estimating oculomotor plant characteristics}
The OPMM describes a system of equations capable of accurately reproducing the characteristics of human eye movements while accounting for the primary components of the oculomotor system.
This is done by estimating the OPC vector of the underlying model of human eye movement.
By default, the OPMM simulates eye movement using either a 9- or 18-parameter model of human eye movement.
These models were developed by Komogortsev \cite{komogortsev20082d} based on earlier work by Bahill \cite{Bahill1980DevelopmentVA}.
Either version can simulate one-dimensional eye movement---either vertical or horizontal.
The OPC vectors for the 9- and 18-parameter models and their default values are described below.

\subsubsection{18-parameter OPMM}
\begin{figure*}
    \centering
    \includegraphics[width=\linewidth]{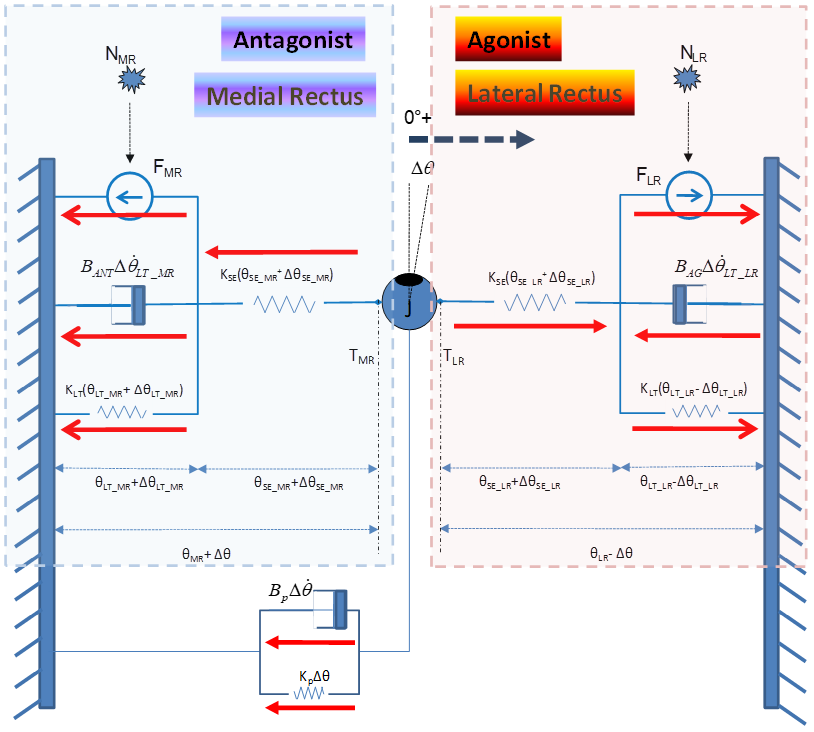}
    \caption{The 18-parameter OPMM employed for positive amplitude saccades. Arrows show the direction of forces for each component. $\Delta\theta$ represents the eye rotation. $J$ is the rotational inertia of the eye globe.}
    \label{fig:opmm_model}
\end{figure*}

The 18-parameter OPC vector includes all major mechanical and neurological components for both the agonist and antagonist muscles.
The relationships between model components are pictured in Figure \ref{fig:opmm_model}.
The model parameters and their default values are shown in Table \ref{tab:opc_18values}.

\begin{table}
    \centering
    \caption{Parameters and their default values for the 18-parameter OPMM. AG and ANT refer to the agonist and antagonist muscles, respectively.}
    \label{tab:opc_18values}
    \setlength{\tabcolsep}{6pt}
    \begin{tabular}{llS[table-format=2.6]}
        \toprule
        Parameter & Shorthand & {Default value} \\ 
        \midrule
        Series elasticity (AG) & $K_{SE\_AG}$ & 2.5 \\ 
        Series elasticity (ANT) & $K_{SE\_ANT}$ & 2.5 \\ 
        Length-tension (AG) & $K_{LT\_AG}$ & 1.2 \\ 
        Length-tension (ANT) & $K_{LT\_ANT}$ & 1.2 \\ 
        Force-velocity (AG) & $B_{AG}$ & 0.046 \\ 
        Force-velocity (ANT) & $B_{ANT}$ & 0.022 \\ 
        Passive viscosity & $B_P$ & 0.06 \\ 
        Tension slope (AG) & $N_{C\_AG}$ & 0.8 \\ 
        Tension slope (ANT) & $N_{C\_ANT}$ & 0.5 \\ 
        Inertial mass & $J$ & 0.000043 \\ 
        Activation time (AG) & $\tau_{AC\_AG}$ & 11.7 \\ 
        Activation time (ANT) & $\tau_{AC\_ANT}$ & 2.4 \\
        Deactivation time (AG) & $\tau_{DE\_AG}$ & 2.0 \\
        Deactivation time (ANT) & $\tau_{DE\_ANT}$ & 1.9 \\
        Tension intercept & $N_{C\_FIX}$ & 14.0 \\
        Neural pulse (AG) & $N_{SAC\_AG}$ & 55 \\
        Neural pulse (ANT) & $N_{SAC\_ANT}$ & 0.5 \\
        Neural pulse width & $PW$ & {$\text{saccade duration} - 6\,ms$} \\
        \bottomrule
    \end{tabular}
\end{table}

\subsubsection{9-parameter OPMM}
The 9-parameter model makes simplifying assumptions about the neuronal control signal, series elasticity, and length-tension relationship.
The model parameters and their default values are shown in Table \ref{tab:opc_9values}.

\begin{table}
    \centering
    \caption{Parameters and their default values for the 9-parameter OPMM. AG and ANT refer to the agonist and antagonist muscles, respectively.}
    \label{tab:opc_9values}
    \setlength{\tabcolsep}{6pt}
    \begin{tabular}{llS[table-format=2.6]}
        \toprule
        Parameter & Shorthand & {Default value} \\
        \midrule
        Series elasticity & $K_{SE}$ & 2.5 \\
        Length-tension & $K_{LT}$ & 1.2 \\
        Force-velocity (AG) & $B_{AG}$ & 0.046 \\
        Force-velocity (ANT) & $B_{ANT}$ & 0.022 \\
        Passive viscosity & $B_P$ & 0.06 \\
        Tension slope (AG) & $N_{C\_AG}$ & 0.8 \\
        Tension slope (ANT) & $N_{C\_ANT}$ & 0.5 \\
        Inertial mass & $J$ & 0.000043 \\
        Tension intercept & $N_{C\_FIX}$ & 14.0 \\
        \bottomrule
    \end{tabular}
\end{table}

\section{Parallel OPMM architecture}
One drawback of the initial OPMM model implementation was parameter estimation speed.
Finding the optimal OPC vector for a given saccade requires an exhaustive search of potential parameter values.
This consumes a large amount of the total simulation time due to the computational demands of OPC estimation.
However, it was observed by the authors that many simulation tasks could be executed in parallel, which could greatly improve overall simulation time.

The initial implementation of the parallel OPMM is written in CUDA, a parallel programming platform developed by NVIDIA to program massively parallel graphics processing units (GPUs) for general purpose computation. Using CUDA, the programmer can develop in a familiar low-level language such as C/C++, then compile the application so it is optimized for a GPU. Once a program has been written in CUDA it can inherit performance improvements in successive generations of CUDA hardware without modification \cite{nvidia2018datasheet,nvidia2018brief}.

\begin{figure}
    \centering
    \includegraphics[height=6cm]{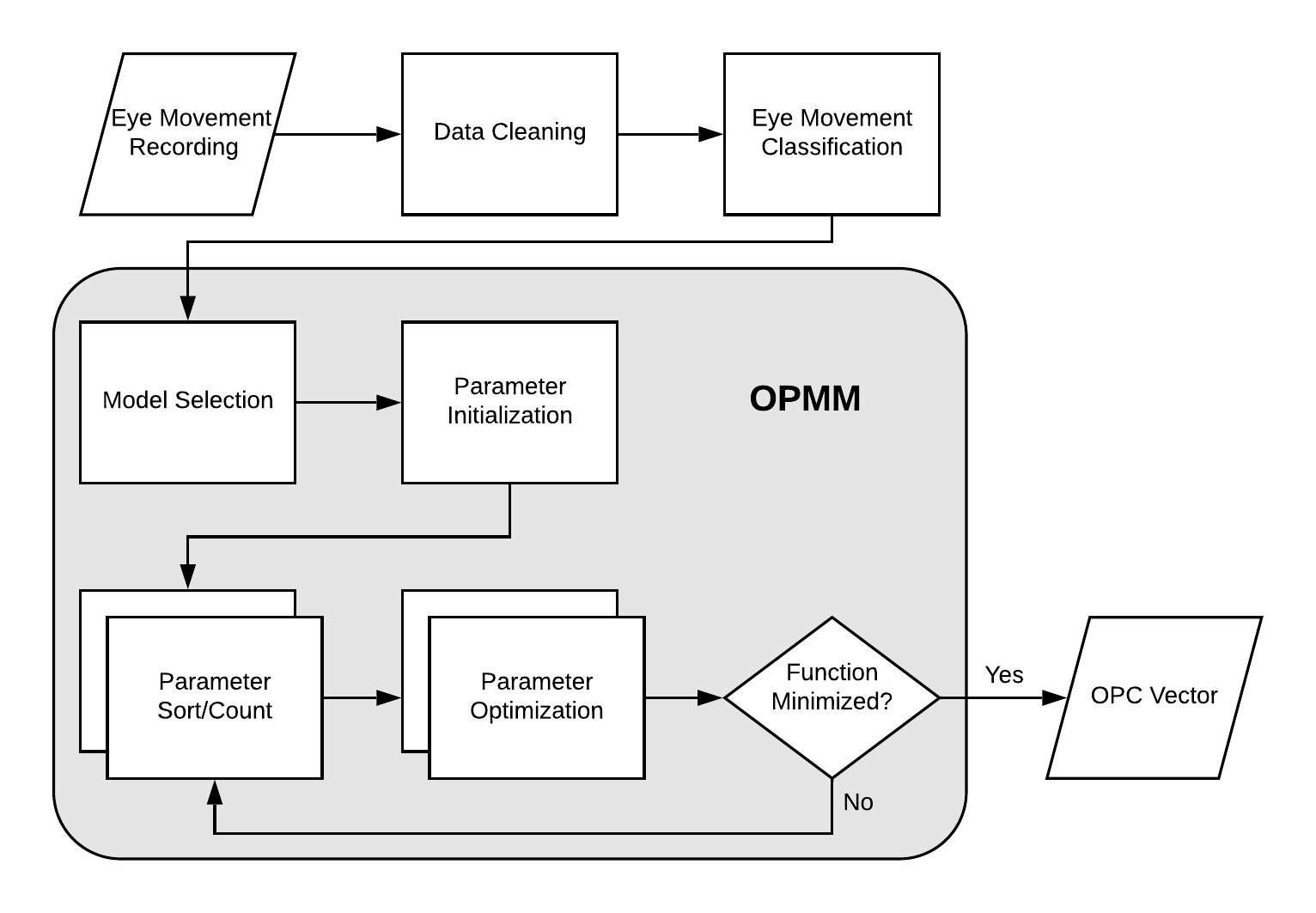}
    \caption{OPMM pipeline}
    \label{fig:opmm_diagram}
\end{figure}

The OPMM pipeline (visualized in Figure \ref{fig:opmm_diagram}) will be discussed in the following sections.


\subsection{OPMM input}
The OPMM uses recorded saccade trajectory data as input. The first three stages of the OPMM analysis pipeline depicted in Figure \ref{fig:opmm_diagram} are concerned with generating high quality input for analysis. These stages occur outside of the OPMM application. Eye movements are typically recorded using video oculography on a commercial eye tracker \cite{duchowski2007eye}. Irregular artifacts are filtered from the recording data during a cleaning phase. These artifacts can be caused by subject movement (e.g. blinking), environmental factors such as ambient moisture, or machine recording error. Saccades are then identified and extracted from the composite eye movement data.  Only the extracted saccades are presented to the OPMM for analysis.

\subsection{Model selection and parameter initialization}
The OPMM has a pluggable architecture and is therefore not tied to a specific model of human eye movement.
It can simulate eye movement using many common models.
The user selects the relevant model at run time.
The OPMM is initialized with the number of parameters required by the chosen model and default values culled from literature.
By default, the OPMM simulates eye movements using an 18-parameter model developed by Komogortsev \cite{komogortsev20082d,komogortsev2008eye}.
It has also been tested with the following alternatives: 

\begin{enumerate}
\item \textit{Komogortsev 2004} \cite{komogortsev2004predictive}: The 9-parameter model is similar to the 18-parameter model but it makes simplifying assumptions about the neuronal control signal, series elasticity, and length-tension relationship. 
\item \textit{Robinson 1965} \cite{robinson1964mechanics}: A 10-parameter model that includes a pulse-step neuronal control signal capable of generating saccades of various amplitudes. 
\item \textit{Enderle 1995 1 and 2} \cite{enderle2010models1}: 19-parameter models. Version 1 recalculates the neuronal control signal at each time step. The control signals are constant in version 2. 
\item \textit{Enderle 2010 1 and 2} \cite{enderle2010models2}: Version 1 has 28 parameters including state switch time constants for the agonist and antagonist muscles and a steady-state parameter for the peak of the agonist pulse. Version 2 has 19 parameters. 
\end{enumerate}

\subsection{Parameter estimation}
Once the model is selected and the model parameters are initialized to default values, the user supplies a vector of eye movement recordings as input.
Then the OPMM uses iterative parameter estimation techniques to find OPC values that minimize the difference between the recorded and simulated saccade trajectories.
Parameter estimation is a type of multi-dimensional search where parameter values are sought that form agreement between a model and data \cite{gershenfeld1999}.
 
The OPMM utilizes a well known parameter estimation algorithm called Nelder-Mead simplex  to minimize the difference between the recorded and simulated saccade trajectory by adjusting the model's default OPC values.
Nelder-Mead is a non-linear, unconstrained multi-dimensional search algorithm \cite{NelderMead1965} that performs an iterative search from an initial guess.
It uses simplex transformations to explore a multi-dimensional topography, searching for values that minimize the difference between the recorded and simulated saccade trajectories. 
Nelder-Mead is not provably optimal \cite{senn2003conversation}. There are situations where it does not converge.

OPMM uses an original parallel implementation of Nelder-Mead based on a serial implementation by Lagarias \cite{lagarias1998convergence}.
The pseudocode in Algorithm \ref{alg:opmm} illustrates the parameter estimation workflow. 
In the parallel version, all of the simplex transformations are calculated in simultaneously. 
The solutions are sorted for accuracy at every iteration.
There are two minimum tolerance thresholds that both must be reached before the minimization routine exits.
First, the maximum coordinate difference between the current best point and the other points in the simplex must be less than or equal to a distance tolerance.
Second, the corresponding difference in function output values must be less than or equal to the absolute difference between the measured and simulated eye movement trajectories.
The Nelder-Mead minimization routine iterates until both of the preceding conditions are true, or until the maximum number of iterations is exceeded.

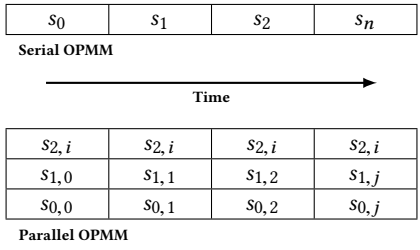
\begin{figure}
    \centering
    \begin{tabular}{| >{\centering}p{1cm} | >{\centering}p{1cm} | >{\centering}p{1cm} | >{\centering}p{1cm} |}
    \hline
    $s_0$ & $s_1$ & $s_2$ & $s_n$ \tabularnewline
    \hline
    \multicolumn{4}{l}{\scriptsize \textbf{Serial OPMM}} \tabularnewline
    \end{tabular}
    
    \vspace{0.2cm}
    
    \begin{tikzpicture}
        \draw[-latex,thick] (0,0) -- node [midway,below] {\scriptsize \textbf{Time}} (4.4,0);
    \end{tikzpicture}
    
    \vspace{0.2cm}
    
    \begin{tabular}{| >{\centering}p{1cm} | >{\centering}p{1cm} | >{\centering}p{1cm} | >{\centering}p{1cm} |}
    \hline
    $s_{2,i}$ & $s_{2,i}$ & $s_{2,i}$ & $s_{2,i}$ \tabularnewline
    \hline
    $s_{1,0}$ & $s_{1,1}$ & $s_{1,2}$ & $s_{1,j}$ \tabularnewline
    \hline
    $s_{0,0}$ & $s_{0,1}$ & $s_{0,2}$ & $s_{0,j}$ \tabularnewline
    \hline
    \multicolumn{4}{l}{\scriptsize \textbf{Parallel OPMM}} \tabularnewline
    \end{tabular}
    \caption{Parallel saccade simulation}
    \label{fig:pic1}
\end{figure}

The parallel OPMM implementation can estimate parameters for thousands of eye movements simultaneously, greatly increasing the rate at which the OPMM can simulate eye movements.
Conducting parameter estimation in parallel reduces the latency of the entire simulation task.
Reducing latency and increasing throughput are both critical considerations for applying the OPMM to large-scale eye movement analysis tasks.
Figure \ref{fig:pic1} depicts serial versus parallel saccade simulation. The parallel version processes multiple saccades at every time step.

\begin{algorithm}
    \SetAlgoLined\SetArgSty{}
    \SetKwInOut{Input}{Input}\SetKwInOut{Output}{Output}
    \SetKwFunction{sort}{sort}
    
    \Input{Saccade trajectory data and initial model parameters}
    \Output{Oculomotor plant characteristic vector that minimizes the error function}
    
    \ForEach{saccade $s$}{
        Spawn CUDA optimization task $t$\;
        Set initial NM-simplex values $vt$ to model values $v$\;
    }
    
    \ForEach{task $t$}{
        \sort{fvt}\;
    }
    
    \While{non-finished tasks present}{
        \ForEach{task $t$}{
            \If{exit criteria for test $t$ is met}{
                Mark $t$ complete\;
            }
        }
        
        \If{all tasks finished}{
            break\;
        }
        
        \ForEach{active task $t$}{
            Prepare reflection points $xr$, expansion points $xe$, outside contraction points $xco$, inside contraction points $xci$, and shrink points $xs$\;
            Perform NM decision step\;
            \sort{fvt}\;
        }
    }
    
    \Return{$vt1$}
    
    \caption{OPMM extraction algorithm}
    \label{alg:opmm}
\end{algorithm}

As mentioned previously, the parallel model was implemented for the CUDA architecture.
CUDA has a heterogeneous computing model.
Data is initialized on the host CPU then transferred to the GPU memory for processing.
The GPU returns the results to the CPU when the processing is complete.
Transferring data between the host and GPU memory introduces latency that slows performance.
Therefore, the parallel OPMM was designed to simultaneously processes all saccades in the input set on the GPU in order to minimize the data transfer overhead.

\subsection{OPMM output}
Figure \ref{fig:output1} shows example OPMM output for simulated saccades using the Komogortsev 18-parameter model.
The output consists of an OPC vector of model parameter values that minimizes the optimization error between the simulated trajectory and the recorded trajectory of each eye movement.
\texttt{SacNo} is the saccade number, \texttt{OptErr} is the optimization error computed by the parallel implementation, \texttt{CPU\_check} is the optimization error computed by a serial implementation for validation purposes, \texttt{SE\_ag} and \texttt{SE\_ant} are the agonist and antagonist series elasticity, \texttt{LT\_ag} and \texttt{LT\_ant} are the agonist and antagonist length-tension, \texttt{PE\_ag} and \texttt{PE\_ant} are the agonist and antagonist tension slope, \texttt{Vis} is the passive viscosity, \texttt{FV\_ag} and \texttt{FV\_ant} are the agonist and antagonist force-velocity, \texttt{Inert} is the interital mass, \texttt{Act\_ag} and \texttt{Act\_ant} are the agonist and antagonist activation time, \texttt{Deact\_ag} and \texttt{Deact\_ant} are the agonist and antagonist deactivation time, \texttt{Step} is the tension intercept, \texttt{H\_ag} and \texttt{H\_ant} are the agonist and antagonist neural pulse height, and \texttt{W} is the neural pulse width.

\begin{figure}
    \centering
    \begin{lstlisting}[frame=single]
SacNo,OptErr,CPU_check,SE_ag,SE_ant,LT_ag,LT_ant,PE_ag,PE_ant,Vis,FV_ag,FV_ant,Inert,Act_ag,Act_ant,Deact_ag,Deact_ant,Step,H_ag,H_ant,W
1,10.701933,10.701933,25.000000,2.035655,1.661605, 0.120000,8.000000,0.050064,0.006063,0.026839, 0.050331,0.000043,5.818619,1.517864,1.526640, 1.518075,6.410131,3.022952,19.199562,14.02288
2,1.115680,1.115681,2.903223,2.599483,0.746196,1.110741, 0.874736,0.513325,0.008178,0.078129,0.022898, 0.000054,15.102122,1.968479,2.562555,1.695454, 2.916430,126.041275,5.574565,7.316801
...
    \end{lstlisting}
    \caption{Example output from OPMM 18-parameter model}
    \label{fig:output1}
\end{figure}

The optimization error is the absolute difference between the recorded and simulated eye movement trajectories.
Optimization error quantifies the goodness of fit between the model and the input.
It is also used to compare accuracy across models when simulating the same recorded input.

\section{Results}
This section of the paper compares the performance of the original serial OPMM algorithm running in MATLAB to the CUDA parallel implementation.
The MATLAB and CUDA versions were compared for speed and accuracy using well known metrics for computer system performance: 

\begin{enumerate}
    \item \textbf{Speedup}: the relative performance of two computer systems solving the same problem.  
    \item \textbf{Accuracy}: the mean absolute difference between the measured and simulated eye movement trajectories.
    \item \textbf{Throughput}: the rate at which each implementation can return OPC vectors for a series of eye movements.
\end{enumerate}

Input data was recorded from 32 subjects.
Subjects' eye movements were recorded as they followed a jumping dot stimulus with $10^{\circ}$ amplitude in the visual field.
Eye movements were captured on an EyeLink 1000 tracker operating at 1000~Hz sampling frequency.
Subjects were seated 70 cm from the display and rested their heads on a chin rest to ensure high eye movement recording accuracy.
The Velocity-Threshold (I-VT) algorithm was used to identify saccades within the recorded eye movement data.
Saccades with amplitudes of less than $4^{\circ}$/s or durations of less than 6 ms were discarded in order to improve result quality.
All saccades were identified and processed offline. 

\subsection{Test environment} 


The OPMM implementations were compared across two test systems tuned for maximum performance: (1) a MATLAB implementation running on a commodity server and (2) a CUDA implementation running on an NVIDIA TU102 GPU.
They are described in Figure \ref{fig:systems}.

\begin{figure}
\begin{lstlisting}[frame=single]
Dell R740 server
-- Intel Xeon Silver 4116 CPU @ 2.10GHz (24)
-- 8 x 16384MB DIMM DDR4 @2666 MHz (128GB)
-- PCI Express Gen 3 (8.0 GT/s)

NVIDIA 2G183 [Tesla T4]
-- CUDA compilation tools, release 10.1, V10.1.105

Red Hat Enterprise Linux server 7.6 (Maipo)
-- Linux kernel 3.10.0-957.10.1.el7
-- GCC 4.8.5 20150623 (Red Hat 4.8.5-36)
\end{lstlisting}

\caption{Test system hardware and software configuration}
\label{fig:systems}
\end{figure}

\subsection{Performance results} 
Table \ref{tab:table-name} compares performance between the original MATLAB implementation and the CUDA version.
Speedup is the first metric for comparison.
It compares the relative run time of the two implementations.
This metric demonstrates each implementations' fitness for simulating saccades in real time.
The CUDA version achieved a peak relative speedup of 19.55 versus the MATLAB implementation, with a runtime of 42.6 seconds versus 832.8 sec processing the same input data set.

\begin{table}
\centering
\caption{\label{tab:table-name}Performance results}
\setlength{\tabcolsep}{6pt}
\begin{tabular}{lcccc}
    \toprule
    Version & Runtime & Speedup & Residual & Throughput \\
    \midrule
    MATLAB & 832.8 sec & 1.0 & 43.8 & 9.22/sec \\
    CUDA & 42.6 sec & 19.55 & 7.1 & 464.61/sec \\
    \bottomrule
\end{tabular}
\end{table}

Accuracy was compared by computing the absolute residual difference between the recorded and simulated trajectories.
A lower absolute residual difference indicates higher fidelity between the recording and the simulation.
The MATLAB version attained an accuracy of 43.8.
The CUDA implementation achieved an accuracy of 7.1.
Because the Nelder-Mead is not proven to converge, both implementations will exit if the maximum number of iterations is not reached within a time boundary. 
Therefore, the accuracy difference between the serial and parallel implementation are likely due to the parallel version exploring the parameter value search space further than the serial version before the time boundary is reached.

The throughput results compare the rate at which the OPMM implementations processed input when performing approximately 60,000 saccade simulations.
Throughput is measured in simulations per second.
This performance metric assesses fitness for large-scale eye movement simulation.
The MATLAB implementation achieved a maximum throughput of 9.22 samples/second.
The CUDA implementation achieved a maximum throughput of 464.61 samples/second.

\section{Conclusion and future work}

The oculomotor plant mathematical model can reproduce human eye movements by estimating the quantitative anatomical characteristics of the subject's oculomotor plant.
It has already been applied to a wide range of use cases.
These include biometric identification, testing eye-gaze guided computer interfaces, and identifying traumatic brain injuries.
However, the slow parameter estimation speed of the original model makes it impractical for real-time and large-scale applications.
This paper introduces a parallel implementation of the OPMM. 
Written in CUDA, the parallel OPMM is capable of processing thousands of eye movement records simultaneously.
It features a pluggable architecture that supports many well-known models of human eye movement with the same parallel search algorithm.

The authors conducted a performance study evaluating the parallel OPMM's suitability for both large-scale and real-time simulations.
Parallel OPMM performance on a GPU was compared to OPMM performance on a commodity processor.
The parallel OPMM running on a GPU outperformed the CPU version both in terms of speed and accuracy when processing approximately 60,000 saccade simulations, reducing the total processing time from 832.8 seconds down to 42.6 seconds (a 19.5x speedup).

These speed increases translate into an overall throughput increase from 9.22 to 464.61 simulations per second.
Additionally, the residual error fell from 43.8 to 7.1 when switching to the parallel version.
Taken together, the results suggest that the parallel implementation can move the OPMM toward real-world and large-scale applications without sacrificing accuracy.

Although the parallel version showed impressive speedup versus the original, further latency reduction is required in order to process input in true real-time, which can be defined as the same amount of time it takes to record the input.
Additional work is also needed to determine how much of the observed performance improvement is inherent to the parallel algorithm versus the hardware platform differences.
Finally, Nelder-Mead simplex is not provably optimal.
Alternative parameter estimation algorithms---such as gradient descent---should be tested for OPMM to avoid situations where Nelder-Mead does not converge.

\begin{acks}
This material is based upon work supported by the National Science Foundation Graduate Research Fellowship under Grant No. DGE-1144466.
The study was also funded by 3 grants to Dr. Komogortsev: (1) National Science Foundation, CNS-1250718 and CNS-1714623, www.NSF.gov; (2) National Institute of Standards and Technology, 60NANB15D325, www.NIST.gov; (3) National Institute of Standards and Technology, 60NANB16D293.
Any opinions, findings, and conclusions or recommendations expressed in this material are those of the author(s) and do not necessarily reflect the views of the National Science Foundation or the National Institute of Standards and Technology.
\end{acks}

\bibliographystyle{ACM-Reference-Format}
\bibliography{parallel-opmm}










\end{document}